\documentclass[prc,floatfix,groupedaddress,nofootinbib,showpacs,amsmath,amssymb,amsfonts,superscriptaddress,widetable]{revtex4-1}
\usepackage{graphicx}
\usepackage{booktabs}
\usepackage{longtable}
\usepackage[T1]{fontenc}
\begin{document}

\title{Effects of high-$j$ orbitals, pairing and deformed neutron shells on upbendings of ground-state bands in neutron-rich even-even isotopes $^{170-184}$Hf}

\author{Xiao-Tao He}%
\email{Corresponding author: hext@nuaa.edu.cn}
 \affiliation{College of Material Science and Technology, Nanjing University of
              Aeronautics and Astronautics, Nanjing 210016, China}

\author{Yang Cao}%
 \affiliation{College of Material Science and Technology, Nanjing University of
              Aeronautics and Astronautics, Nanjing 210016, China}
      
\author{Xiao-Ling Gan}%
 \affiliation{College of Material Science and Technology, Nanjing University of
              Aeronautics and Astronautics, Nanjing 210016, China}

\date{\today}

\begin{abstract}
The ground-state bands (GSBs) in the even-even hafnium isotopes $^{170-184}$Hf are investigated by using the cranked shell model (CSM) with pairing correlations treated by the particle-number conserving (PNC) method. The experimental kinematic moments of inertia are reproduced very well by theoretical calculations. The second upbending of the GSB at high frequency $\hbar\omega\approx0.5$ MeV observed (predicted) in $^{172}$Hf ($^{170,174-178}$Hf) attributes to the sudden alignments of the proton high-$j$ orbitals $\pi1i_{13/2}$ $(1/2^{+}[660])$, $\pi1h_{9/2}$ $(1/2^{-}[541])$ and orbital $\pi1h_{11/2}$ $(7/2^{-}[523])$. The first upbendings of GSBs at low frequency $\hbar\omega=0.2-0.3$ MeV in $^{170-178}$Hf, which locate below the deformed neutron shell $N=108$, attribute to the alignment of the neutron orbital $\nu1i_{13/2}$. For the heavier even-even isotopes $^{180-184}$Hf, compared to the lighter isotopes, the first band-crossing is delayed to the high frequency due to the existence of the deformed shells $N=108,116$. The upbendings of GSBs in $^{180-184}$Hf are predicted to occur at $\hbar\omega\approx0.5$MeV, which come from the sharp raise of the simultaneous alignments of both proton $\pi1i_{13/2}$, $\pi1h_{9/2}$ and neutron $\nu2g_{9/2}$ orbitals. The pairing correlation plays a very important role in the rotational properties of GSBs in even-even isotopes $^{180-184}$Hf. Its effects on upbendings and band-crossing frequencies are investigated. 
\end{abstract}

%\pacs{21.10.Re, 23.20.Lv, 21.60.Cs, 27.90.+b}%

%U

\maketitle

\section{Introduction}
Hafnium isotopes are well known as nuclei with long-lived high-K isomers, like the $K^{\pi}=16^{+}$ isomer with half life $T_{1/2}=31$ yr in $^{178}$Hf~\cite{MullinsS1997_PLB393}. Here $K$ is a quantum number representing the projection of the total nuclear spin along the symmetry axis of the nucleus. A $K$ isomer appears only in axially symmetric deformed nuclei well away from the closed shell~\cite{WalkerP1999_N399,DracoulisG2016_RoPiP79}. Recent years, with development of experimental techniques, the study of the well deformed rare-earth nuclei moves away from the $\beta$-stable line towards the neutron-rich region. The neutron-rich hafnium isotopes locate in the upper half of the closed shell of protons $(Z = 50, 82)$ and neutrons $(N = 82, 126)$. In this region, it is very interesting that the multi-particle excitation is expected to be increasingly favored~\cite{WalkerP1999_N399, AabergS1978_NPA306, Ngijoi-YogoE2007_PRC75, OiM2001_PLB505}, and a shape transition from prolate to oblate is predicted along the yrast line at the high spin~\cite{HiltonR1979_PRL43,XuF2000_PRC62,TandelU2008_PRL101,SarrigurenP2008_PRC77,OiM2001_PLB505}. In addition to the exotic nuclear structure itself, the knowledge of neutron-rich nuclei is essential to understand nucleosynthesis and the origin of heavy nuclei in the Universe. The experimental data also provide a good testing ground for the present nuclear theories in the neutron-rich rare-earth region.

Valuable structural information is embedded in the high-spin states of hafnium isotopes. Like in $^{172}$Hf, the highest state is established at spin $44\hbar$~\cite{nndc,CullenD1995_PRC52}, and there are two upbendings appearing at the frequency $\hbar\omega\approx0.2$ and $0.5$ MeV of the GSB, respectively. In the heavier neutron-rich isotopes $^{180,182}$Hf, according to the experimental data, there is no obvious nucleon alignment was observed in the GSB up to $\hbar\omega=0.43$ MeV~\cite{Ngijoi-YogoE2007_PRC75}. Compared to the lighter hafnium isotopes, there is a significant delay of the band-crossing. $^{184}$Hf is the heaviest hafnium isotope in which the rotational spectrum were observed up to now~\cite{nndc, KondevF2015_ADaNDT103_104, KrumbholzK1995_ZA351, RykaczewskiK1989_NPA499}. It might have similar rotational character with $^{180,182}$Hf. The different behavior of the rotational GSBs between the lighter and heavier hafnium isotopes is a demonstration of the deformed shell structure along the isotope chain.  

Theoretically, besides studies focused on the high-$K$ isomers~\cite{DracoulisG2016_RoPiP79,WalkerP2001_HI135,SunY2004_PLB589,XuF2000_PRC62,ZengJ2002_PRC65,ZhangZ2009_PRC80}, several theoretical investigations have been performed on the high angular moment states of the neutron-rich hafnium isotopes. Shape coexistence and the high-spin states near the yrast line in $^{182}$Hf were studied by tilted-axis-cranked Hartree-Fock-Bogoliubov calculations~\cite{OiM2001_PLB505}. A comprehensive study on the low-lying states of neutron-rich hafnium isotopes across the $N = 126$ shell was performed by a multireference covariant density functional theory~\cite{WuX2019_PRC99}. The study of collective rotations in $^{172-178}$Hf were carried out by configuration-constrained calculations~\cite{,LiangW2015_PRC92}. Nuclear pairing correlations in $^{170-182}$Hf were investigated by PNC-CSM~\cite{ZengJ1994_PRC50}. Ground-state properties of $^{156-168}$Hf were studied by the Nilsson mean-field plus extended-pairing model~\cite{GuanX2015_PRC92}. Band-crossing frequencies at high rotational frequencies were investigated in the rare-earth nuclei around hafnium~\cite{MillerS2019_PRC100_14302}.

Systematical investigation of the high-spin states in a comparatively long hafnium isotope chain is absent, particularly ones that include the very neutron-rich hafnium isotopes $^{180,182,184}$Hf. Such investigation is helpful to understand better of the nuclear structure of exotic nuclei in the neutron-rich rare-earth region, like nuclear shape, deformed shells, high-$j$ intruder orbitals, band-crossing, pairing and so on. In the present work, the GSBs in the neutron-rich even-even isotopes $^{170-184}$Hf are studied by using the cranked shell model with pairing treated by the particle-number conserving method. The microscopic mechanism of the upbending which caused by the alignment of the paired nucleon occupying high-$j$ orbitals is explained, including the upbending at the low frequency in the lighter even-even isotopes $^{170-178}$Hf, the second upbending at the high frequency in $^{172}$Hf, the absence of alignment at low frequency and upbendings at high frequency in the heavier even-even isotopes $^{180-184}$Hf. Based on these studies, the deformation, deformed neutron shell, the high-$j$ intruder orbitals and pairing correlation are discussed in the neutron-rich rare-earth region.    

\section{Theoretical Framework}\label{sec:2}

The cranked shell model Hamiltonian in the rotating frame is,
\begin{eqnarray}
 H_\mathrm{CSM}= H_{\rm SP}-\omega J_x + H_\mathrm{P},
\end{eqnarray}
where $H_{\rm SP}=\sum_{\xi}(h_\mathrm{Nil})_{\xi}$ is the single-particle part with $h_{\mathrm{Nil}}$ being the Nilsson Hamiltonian. $\xi$ ($\eta$) is the eigen state of the Hamiltonian $h_{\xi(\eta)}$ and $\bar{\xi}$ ($\bar{\eta}$) denotes its time-reversed state. $-\omega J_x$ is the Coriolis interaction with the rotational frequency $\omega$ about the $x$ axis (perpendicular to the nuclear symmetry z axis). The cranked single-particle Hamiltonian $h_{0}(\omega)=h_{\xi}-\omega j_{x}$ is diagonalized to obtain the cranked Nilsson levels $\epsilon_{\mu}$ and cranked state $|\mu\rangle$.     
The pairing $H_{\text{P}}=H_{\text{P}}(0)+H_{\text{P}}(2)$ includes monopole- and quadrupole-pairing correlations,   
\begin{eqnarray}
 \ H_{\text{P}}(0)
 &=&-G_{0}\sum_{\xi \eta }
 a_{\xi }^{\dagger}a_{\overline{\xi }}^{\dagger }a_{\overline{\eta }}a_{\eta }\ ,
 \\
 H_{\text{P}}(2)
 &=&-G_{2}\sum_{\xi \eta } q_{2}(\xi) q_{2}(\eta)
 a_{\xi}^{\dagger } a_{\overline{\xi}}^{\dagger}
 a_{\overline{\eta}}a_{\eta}\ ,
 \label{eq:Hp}
\end{eqnarray}
where $q_{2}(\xi) = \sqrt{{16\pi}/{5}}\langle \xi |r^{2}Y_{20} | \xi \rangle$ is the diagonal element of the stretched quadrupole operator. $G_{0}$ and $G_{2}$ (in units of MeV) are the monopole and quadrupole effective pairing strengths, respectively, which are connected with the dimension of the truncated CMPC space. In the present calculations, the CMPC space is constructed in proton $N = 4, 5, 6$ and neutron $N=5, 6$ major shells. The dimensions of the CMPC space are about 800 for protons and 900 for neutrons. The pairing strengths are determined by the odd-even differences in moment of inertia of the hafnium isotopes. The monopole effective pairing strengths are both equal to $G_{0}=0.3$ for protons and neutrons. The stability of the PNC calculation results against the change of the dimension of the CMPC space has been investigated in Refs.~\cite{ZengJ1994_PRC50,LiuS2002_PRC66}.

The cranked shell model Hamiltonian is expressed in the cranked basis as,
\begin{eqnarray}
 H_\mathrm{CSM}= \sum_{\mu}\epsilon_{\mu}b_{\mu}^{\dagger}b_{\mu} 
 -G_0\sum_{\mu\mu{\prime}\nu\nu{\prime}}f^{\ast}_{\mu\mu{\prime}}f_{\nu{\prime}\nu}b_{\mu+}^{\dagger}b_{\mu{\prime}-}^{\dagger}b_{\nu-}b_{\nu{\prime}+}
-G_2\sum_{\mu\mu{\prime}\nu\nu{\prime}}g^{\ast}_{\mu\mu{\prime}}g_{\nu{\prime}\nu}
b_{\mu+}^{\dagger}b_{\mu{\prime}-}^{\dagger}b_{\nu-}b_{\nu{\prime}+},
\label{Eq:H_cranked}
\end{eqnarray}
where $b_{\mu}^{\dagger}$ is the real particle creation operator of the cranked state $|\mu\rangle$. $H_\mathrm{CSM}$ is diagonalized in a sufficiently large cranked many-particle configuration (CMPC) space to obtain the yrast and low-lying eigenstates,
\begin{eqnarray}
 | \psi \rangle = \sum_{i} C_i | i \rangle,
\end{eqnarray}
where $| i \rangle$ denotes an occupation of particles in cranked orbitals and $C_i$ is the corresponding probability amplitude.

The angular momentum alignment $\left\langle J_{x} \right\rangle$ of the state $\left\vert \psi \right\rangle$ is given by
\begin{equation}
 \left\langle \psi \right| J_{x}
 \left| \psi \right\rangle
 = \sum_{i}\left|C_{i}\right| ^{2}
   \left\langle i\right| J_{x}\left| i\right\rangle
 + 2\sum_{i<j}C_{i}^{\ast }C_{j}
   \left\langle i\right| J_{x}\left| j\right\rangle\ .
 \label{eq:Jx}
\end{equation}
The kinematic moment of inertia reads
\begin{equation}
 J^{(1)} = \frac{1}{\omega} \langle\psi | J_x | \psi \rangle \ .
\end{equation}
Since $J_x$ is an one-body operator, the matrix element $\langle i | J_x | j \rangle$ for $i\neq j$ does not vanish only when $|i\rangle$ and $|j\rangle$ differ by one particle occupation.
After a certain permutation of creation operators, $|i\rangle$ and $|j\rangle$ can be recast into
\begin{equation}
 |i\rangle=(-1)^{M_{i\mu}}|\mu\cdots \rangle \ , \qquad
 |j\rangle=(-1)^{M_{j\nu}}|\nu\cdots \rangle \ ,
\end{equation}
where the ellipsis stands for the same particle occupation, and $(-1)^{M_{i\mu}}=\pm1$, $(-1)^{M_{j\nu}}=\pm1$ according to whether the permutation is even or odd. Then the kinematic moment of inertia of $|\psi\rangle$ can be written as
\begin{equation}
 J^{(1)} = \sum_{\mu}j^{(1)}_{\mu}+\sum_{\mu<\nu}j^{(1)}_{\mu\nu} \ ,
 \label{eq:J(1)}
\end{equation}
where the direct and interference terms are,
\begin{eqnarray}
 j^{(1)}_{\mu}   & = & \frac{n_\mu}{\omega}\langle\mu|j_{x}|\mu\rangle  \ ,
 \nonumber \\
 j^{(1)}_{\mu\nu}& = & \frac{2}{\omega} \langle\mu|j_{x}|\nu\rangle
                       \sum_{i<j} (-1)^{M_{i\mu}+M_{j\nu}} C_{i}C_{j} \ ,
  \qquad (\mu\neq\nu) \ ,
 \nonumber
\end{eqnarray}
with
\begin{equation}
 n_{\mu} = \sum_{i}|C_{i}|^{2}P_{i\mu} \ ,
\end{equation}
being the occupation probability of the cranked orbital $|\mu\rangle$ with $P_{i\mu}=1$ if $|\mu\rangle$ being occupied in $|i\rangle$ and $P_{i\mu}=0$ otherwise. The total particle number $N = \sum_\mu n_\mu$ is exactly conserved. 

Since the total Hamiltonian is diagonalized directly in a truncated Fock space, the sufficiently accurate solutions can be obtained in a comparatively small diagonalization space for the yrast and low-lying excited states~\cite{WuC1989_PRC39}. Like in the standard shell model, the particle-number keeps conserved and the Pauli blocking effect is taken into account strictly. The details of the PNC-CSM method can be found in Refs.~\cite{ZengJ1983_NPA405,WuC1989_PRC39,ZengJ1994_PRC50,XinX2000_PRC62,HeX2018_PRC98}. 

\section{Results and discussions}\label{sec:parameters}
\subsection{Cranked Nilsson single-particle levels}

%%%%%%%%%%%%%%%%%%%%%%%%%%%%%%%%%%%%%%%%%%%%%
\begin{table*}
\centering
\caption{Deformation parameters $\varepsilon_2$ and $\varepsilon_4$ used in the present PNC-CSM
calculations for even-even $^{170-184}$Hf isotopes.}
\begin{ruledtabular}
\begin{tabular}{ccccccccc}
nuclei  & $^{170}$Hf	& $^{172}$Hf& 	$^{174}$Hf& 	$^{176}$Hf& 	$^{178}$Hf& 	$^{180}$Hf& 	$^{182}$Hf& 	$^{184}$Hf \\
	 \hline
$\varepsilon_{2}$       & 0.245 	& 0.254	& 0.253	& 0.248	& 0.238	& 0.235	& 0.228	& 0.214	\\
$\varepsilon_{4}$       & 0.014 	& 0.023	& 0.034	& 0.043	& 0.056	& 0.062	& 0.069	& 0.065	
\end{tabular}
\end{ruledtabular}
\label{tab:parameters_defs}
\end{table*} 
%%%%%%%%%%%%%%%%%%%%%%%%%%%%%%%%%%%%%%%%%%%%%

%%%%%%%%%%%%%%%%%%%%%%%%%%%%%%%%%%%%%%%%%%%%%%%%%%%%%
\begin{figure}[ht] 
\centering
    \includegraphics[width=10cm]{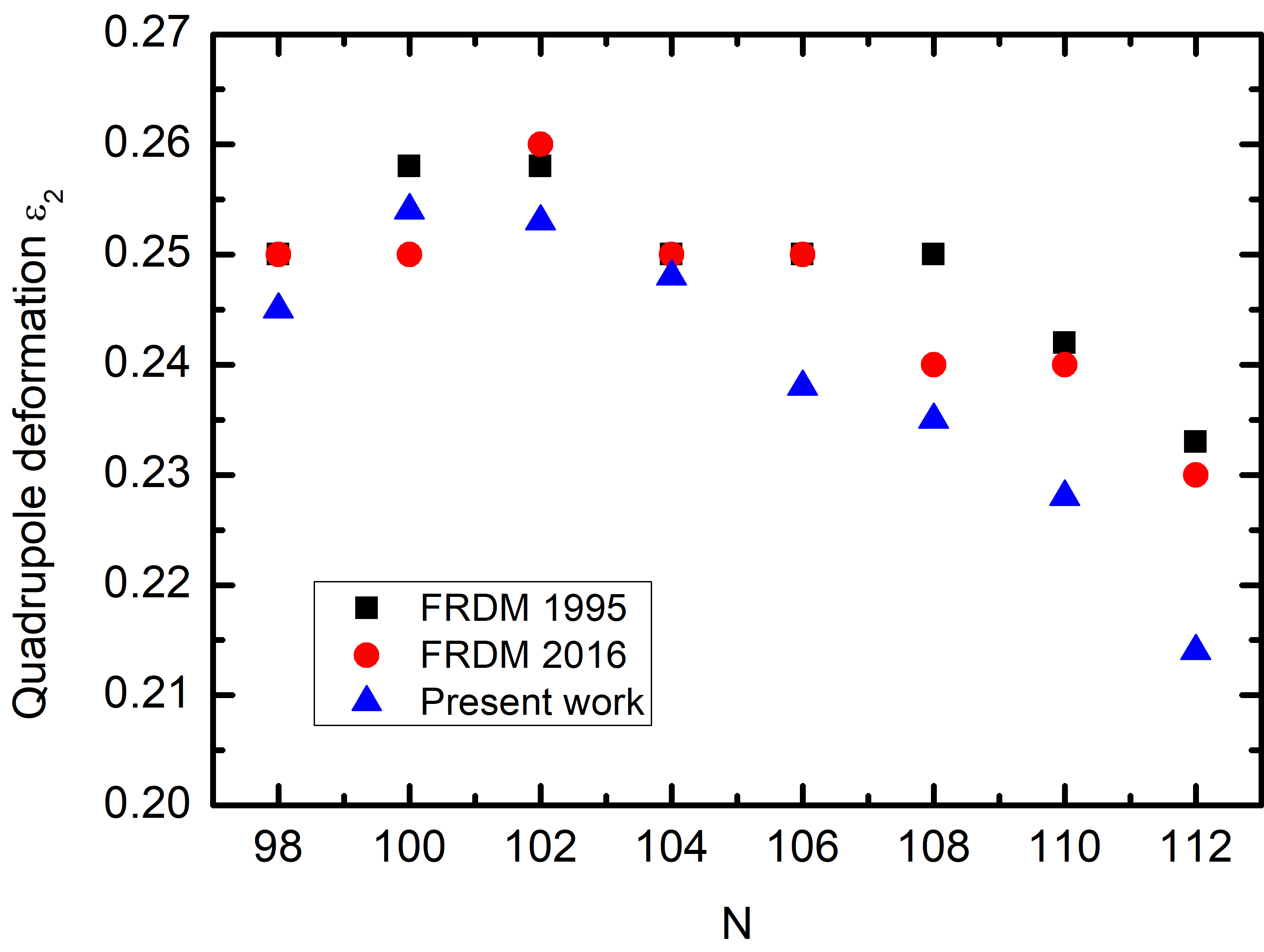}
      \caption{(Color online) Comparison of the quadrupole deformations used in present work (blue triangles) with those predicted by the finite range droplet model in 1995~\cite{MoellerP1995_ADaNDT59} (black squares) and in 2016~\cite{MoellerP2016_ADNDT109-110} (red solid circles) for the even-even isotopes $^{170-184}$Hf.}
\label{fig:Fig1}
\end{figure}
%%%%%%%%%%%%%%%%%%%%%%%%%%%%%%%%%%%%%%%%%%%%%%%%%%%%%

The Nilsson parameters $(\kappa,\mu)$ are taken from Ref.~\cite{NilssonS1969_NPA131}. The deformation parameters $(\varepsilon_{2}, \varepsilon_{4})$ used in the present calculations are listed in Table.~\ref{tab:parameters_defs}, which are chosen by referring to the table of M\"{o}ller \textit{et al}. in 1995~\cite{MoellerP1995_ADaNDT59} and 2016~\cite{MoellerP2016_ADNDT109-110}. Figure~\ref{fig:Fig1} shows the quadrupole deformation given in Table~\ref{tab:parameters_defs} (blue triangles), which are compared with those predicted by the finite range droplet model (FRDM) in 1995~\cite{MoellerP1995_ADaNDT59} (black squares) and in 2016~\cite{MoellerP2016_ADNDT109-110} (red solid circles). One can see that the deformation reaches maximum at $N=100(102)$, $N=102$ and $N=100$ in the prediction from FRDM in 1995, in 2016 and in the input values of the present work, respectively. These agree with the peak of the deformation beyond the midshell, i.e. near $N=101$, which appears common to both the theoretical and measured deformation~\cite{LevinsJ1999_PRL82}. The discrepancy appears mainly after the peak. The deformation decreases as the neutron number increases for the heavier isotopes, but with smaller values and in a smoother way in the present work. This agrees with the study of the measured charge radii in the isotope $^{170-182}$Hf that the nuclear deformations are lower (at a level less than $5\%$ of $\beta_{2}$) than the droplet model prediction~\cite{LevinsJ1999_PRL82}. In the present work, smaller deformation in heavier neutron-rich isotopes are requested by the rotational character at high-spin along the hafnium isotope chain, which will be discussed later. 
 
%%%%%%%%%%%%%%%%%%%%%%%%%%%%%%%%%%%%%%%%%%%%%%%%%%%%%
\begin{figure}[ht] 
\centering
    \includegraphics[width=16cm]{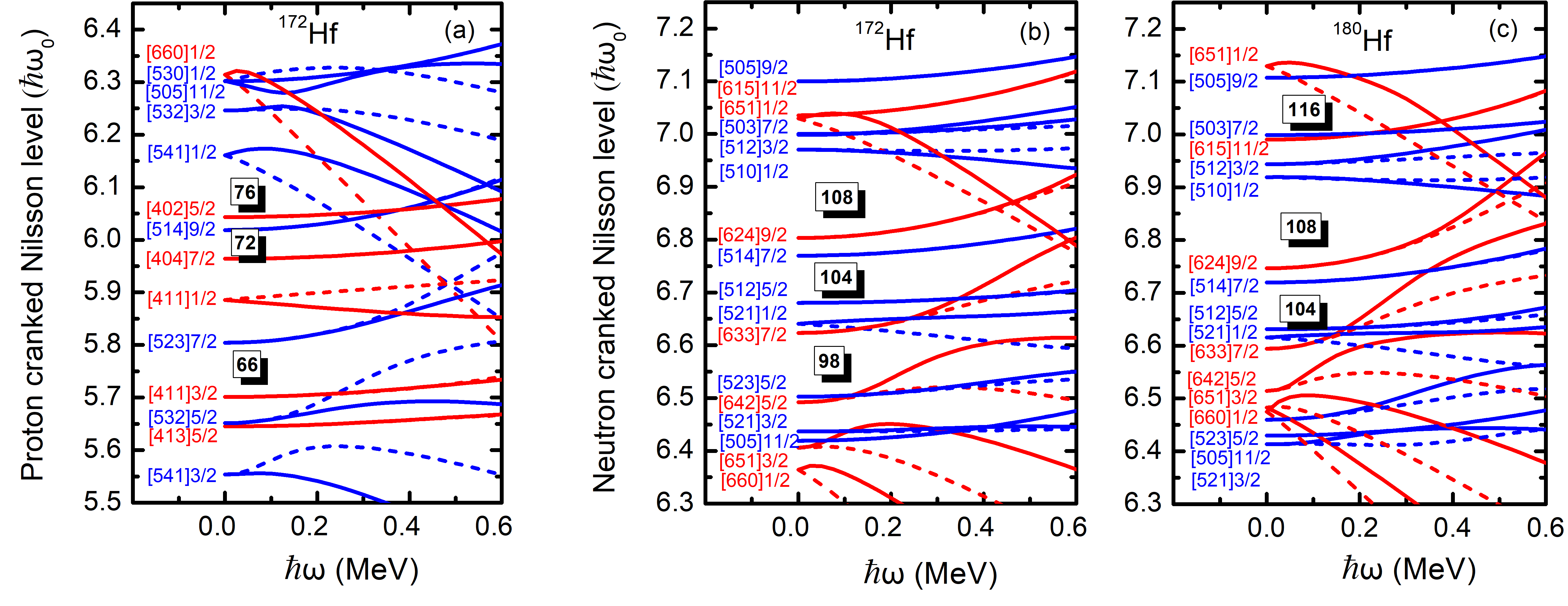}
      \caption{(Color online) The cranked Nilsson levels near the Fermi surface (a) of $^{172}$Hf for protons, (b) of $^{172}$Hf for neutrons, and (c) of $^{180}$Hf for neutrons. The positive (negative) parity levels are denoted by red (blue) lines. The signature $\alpha=+1/2$ $(\alpha= −1/2)$ levels are denoted by solid (dashed) lines.}
\label{fig:Fig2}
\end{figure}
%%%%%%%%%%%%%%%%%%%%%%%%%%%%%%%%%%%%%%%%%%%%%%%%%%%%%

%%%%%%%%%%%%%%%%%%%%%%%%%%%%%%%%%%%%%%%%%%%%%%%%%%%%%
%\begin{figure}[ht] 
%\centering
%    \includegraphics[width=18cm]{Fig3.png}
%      \caption{(Color online) The same as Fig.~\ref{fig:Fig2}, but for the cranked neutron Nilsson orbitals near the Fermi surface of $^{172,178,182}$Hf.}
%\label{fig:Fig3}
%\end{figure}
%%%%%%%%%%%%%%%%%%%%%%%%%%%%%%%%%%%%%%%%%%%%%%%%%%%%%

The proton Fermi surfaces of the whole hafnium isotopes are the same. The proton single-particle level structure near the Fermi surface are similar and only that of $^{172}$Hf is presented. Figure~\ref{fig:Fig2} (a) shows the cranked proton Nilsson levels near the Fermi surface of $^{172}$Hf. The positive (negative) parity levels are denoted by red (blue) lines. The signature $\alpha=+1/2$ $(\alpha=-1/2)$ levels are denoted by solid (dashed) lines. The deformed proton shells are shown at $Z=66,68,70,76$. The high-j intruder orbitals $\pi1/2^{+}[660]$ and $\pi1/2^{-}[541]$ go down rapidly with increasing rotational frequency and cross with $\pi7/2^{+}[404]$ at $\hbar\omega\approx0.4$ MeV around the proton Fermi surface. These orbitals would bring significant influence on the rotational properties of hafnium isotopes.

For neutrons, the Fermi surfaces of $^{170-184}$Hf are different. The single-particle level structure, which is influenced by the deformation,  effects the nuclear rotational properties strongly. In Fig.~\ref{fig:Fig2} (b) and (c), the neutron Nilsson levels near the Fermi surface of $^{172}$Hf (that of $^{170,174,176}$Hf are similar) and $^{180}$Hf (that of $^{178,182,184}$Hf are similar) are displayed, respectively. One can see that the high-$j$ orbitals around the Fermi surface are $\nu 1i_{13/2}$ (including $\nu5/2^{+}[642]$,  $\nu7/2^{+}[633]$, $\nu9/2^{+}[624]$, $\nu11/2^{+}[615]$ deformed levels) and $\nu2g_{9/2}$ (including $\nu1/2^{+}[651]$ level) orbitals. When we compare the neutron Nilsson levels for $^{172}$Hf and $^{180}$Hf, it is found that the deformed neutron shells are $N=98,104,108$ for $^{172}$Hf (with larger deformation $\varepsilon_{2}=0.254$) and $N=104,108,116$ for $^{180}$Hf (with smaller deformation $\varepsilon_{2}=0.235$), respectively. The $N=98$ deformed shell becomes much smaller and the one at $N=116$ emerges for $^{180}$Hf, and meanwhile, the high-$j$ low-$\Omega$ intruder orbital $\nu1/2^{+}[651]$ moves upward and leaves the Fermi surface for $^{180}$Hf. The different behavior of the GSBs for $^{170-184}$Hf isotopes can be understood based on such a single-particle levels structure.  

\subsection{Kinematic moments of inertia}
\label{sec:MoI}

%%%%%%%%%%%%%%%%%%%%%%%%%%%%%%%%%%%%%%%%%%%%%%%%%%%%%
\begin{figure}[ht] 
\centering
    \includegraphics[width=16cm]{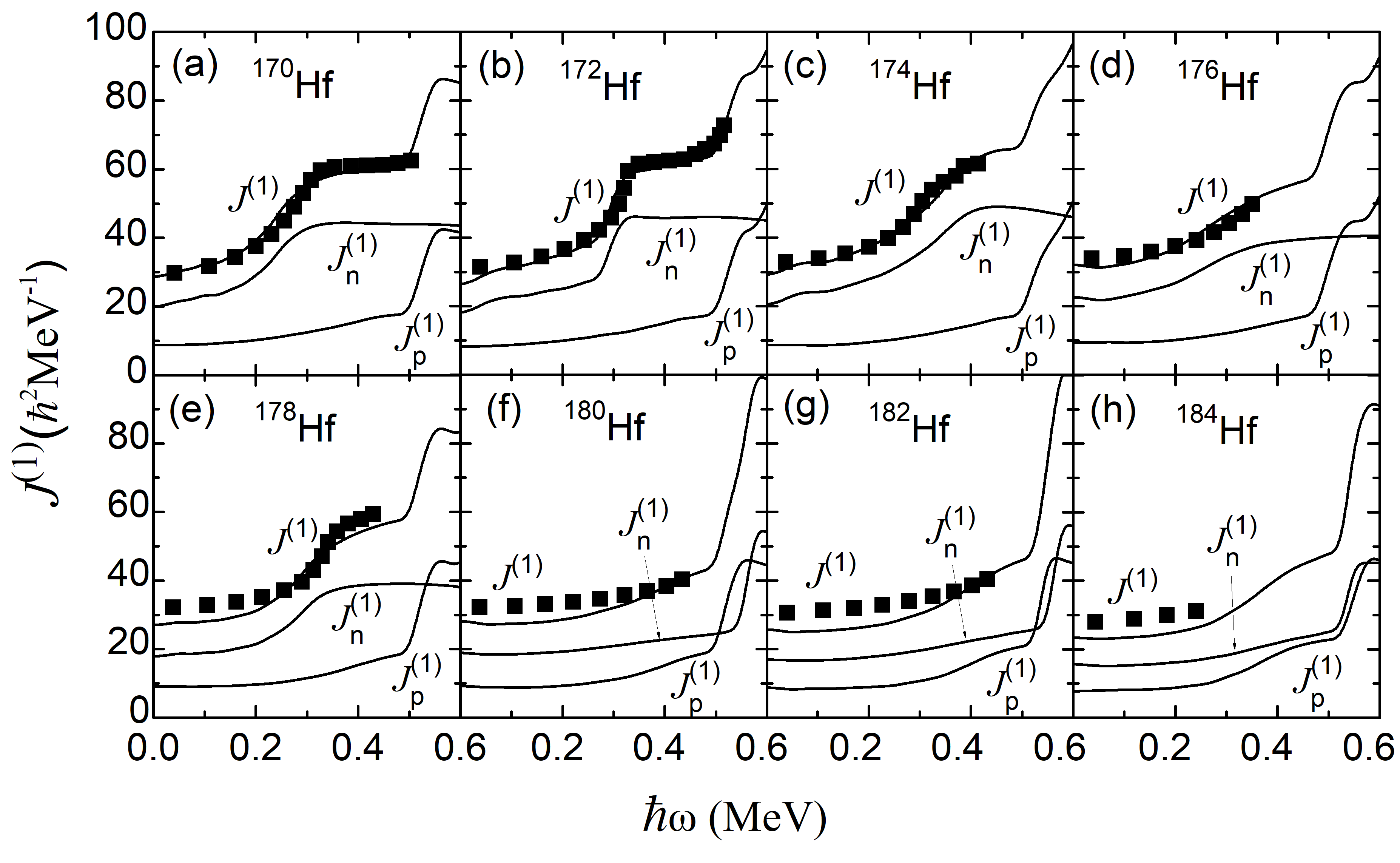}
      \caption{The ground-state bands in the even-even isotopes $^{170-184}$Hf. The experimental kinematic moment of inertia $J^{(1)}$ is denoted by solid square and the theoretical result is plotted by solid line. $J^{(1)}_p$ and $J^{(1)}_n$ are the contributions from protons and neutrons, respectively.}
\label{fig:Fig3}
\end{figure}
%%%%%%%%%%%%%%%%%%%%%%%%%%%%%%%%%%%%%%%%%%%%%%%%%%%%%

The kinematic moments of inertia $J^{(1)}$ of the GSBs in the even-even isotopes $^{170-184}$Hf are shown in Fig.~\ref{fig:Fig3}. The experimental data, which are taken from Ref.~\cite{nndc}, are denoted by solid squares and theoretical results are denoted by solid lines. $J^{(1)}$ shows upbendings at the low frequency region around $\hbar\omega=0.2-0.3$ MeV in the lighter isotopes $^{170-178}$Hf while it increases smoothly during the whole observed frequency region in the heavier isotopes $^{180-184}$Hf. In addition, a second upbending was observed at $\hbar\omega\approx0.5$ MeV of the GSB in $^{172}$Hf. The theoretical calculations reproduce the existent experimental data very well and predict upbendings at $\hbar\omega\approx0.5$ MeV for all the isotopes studied in the present work. 
 
The calculated contributions to $J^{(1)}$ from protons and neutrons are given by $J^{(1)}_p$ and $J^{(1)}_n$, respectively. Since the proton Fermi surface is same, the contributions from the proton are very similar for all the hafnium isotopes. The $J^{(1)}_p$ contributes mainly to the upbending of $J^{(1)}$ at the high frequency $\hbar\omega\approx0.5$ MeV, that is the second upbending for $^{170-178}$Hf while is the first one for $^{180-184}$Hf. Since the neutron Fermi surface are different for these isotopes, the contributions from neutrons depend strongly on the detailed single-particle level structure, such as the deformed shells and high-$j$ intruder orbitals. The first upbending, at low frequency for $^{170-178}$Hf and high frequency for $^{180-184}$Hf, are mainly due to the contribution of neutrons $J^{(1)}_n$. Thus, the different behavior of the first upbending in the hafnium isotopes  $^{170-184}$Hf demonstrates the deformed neutron single-particle structure in the rare-earth neutron-rich region. 

\subsection{Effect of proton high-$j$ orbitals on upbendings at high rotational frequencies}

%%%%%%%%%%%%%%%%%%%%%%%%%%%%%%%%%%%%%%%%%%%%%%%%%%%%%
\begin{figure}[ht] 
\centering
    \includegraphics[width=6cm]{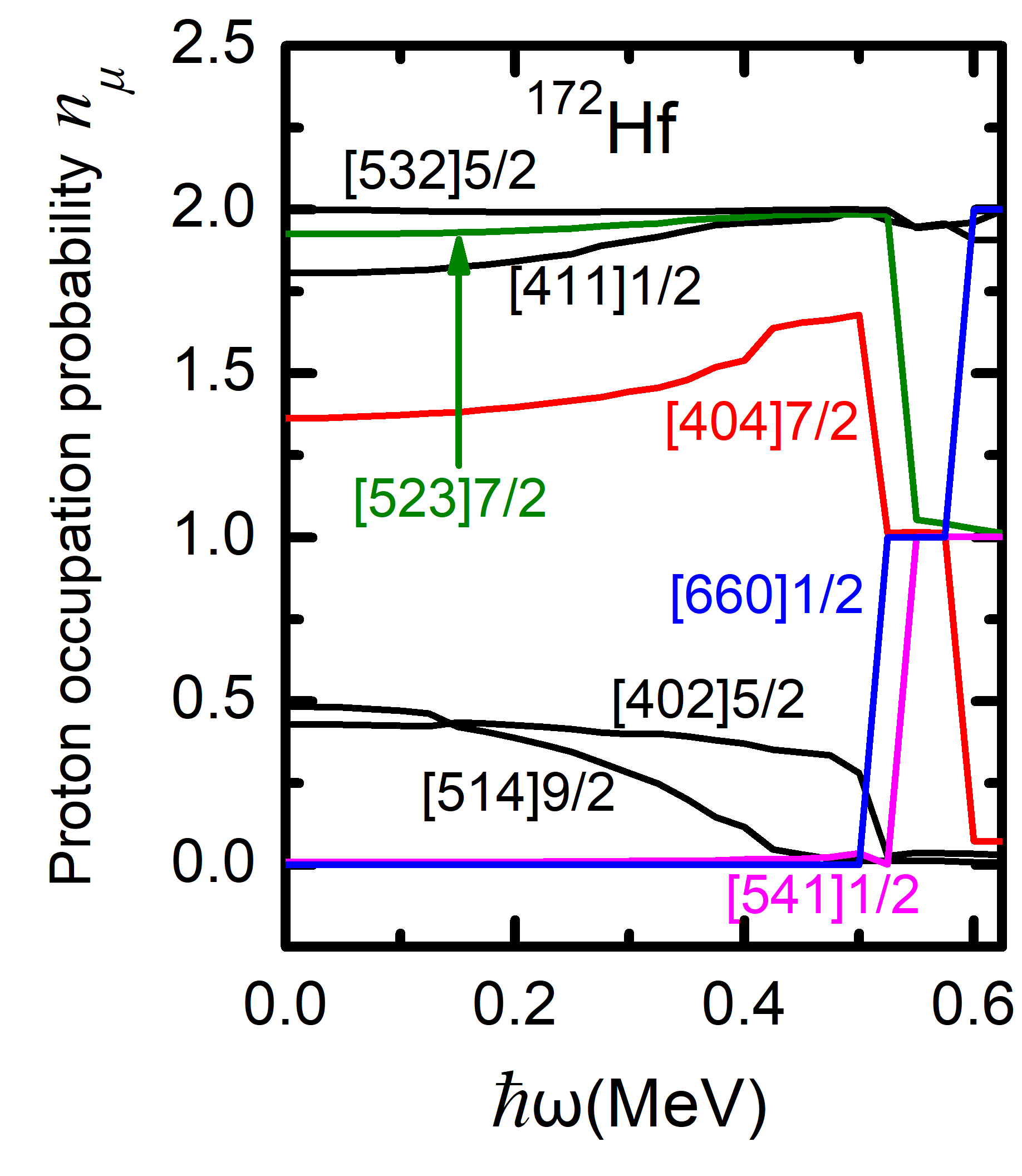}
      \caption{(Color online) Occupation probabilities $n_{\mu}$ of each proton orbital $\mu$ (including both $\alpha=\pm1/2$) near the Fermi surface for the GSB in $^{172}$Hf. The Nilsson levels far above ($n_{\mu}\approx0$) and far below ($n_{\mu}\approx2$) the Fermi surface are not labelled.}
\label{fig:Fig4}
\end{figure}
%%%%%%%%%%%%%%%%%%%%%%%%%%%%%%%%%%%%%%%%%%%%%%%%%%%%%

%%%%%%%%%%%%%%%%%%%%%%%%%%%%%%%%%%%%%%%%%%%%%%%%%%%%%
\begin{figure}[ht] 
\centering
    \includegraphics[width=13cm]{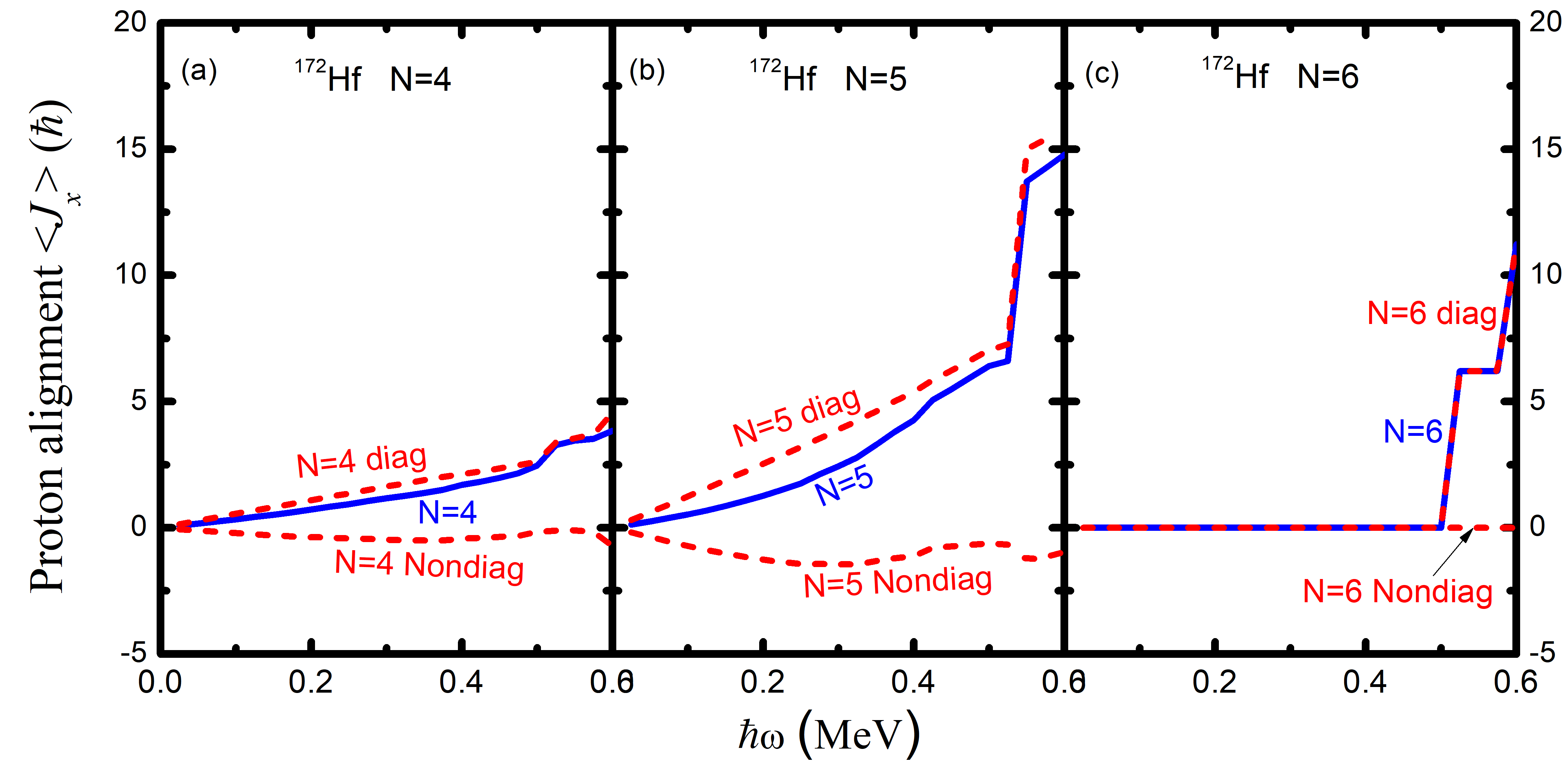}
      \caption{(Color online) Contributions to the angular momentum alignment $\langle J_x \rangle$ from proton $N=4,5,6$ major shells for the GSB in $^{172}$Hf. The diagonal $\sum_{\mu}j^{(1)}_{\mu}$ and off-diagonal $\sum_{\mu<\nu} j^{(1)}_{\mu\nu}$ parts in Eq.~\ref{eq:J(1)} are shown by red dashed lines.}
\label{fig:Fig5}
\end{figure}
%%%%%%%%%%%%%%%%%%%%%%%%%%%%%%%%%%%%%%%%%%%%%%%%%%%%%

%%%%%%%%%%%%%%%%%%%%%%%%%%%%%%%%%%%%%%%%%%%%%%%%%%%%%
\begin{figure}[ht] 
\centering
    \includegraphics[width=13cm]{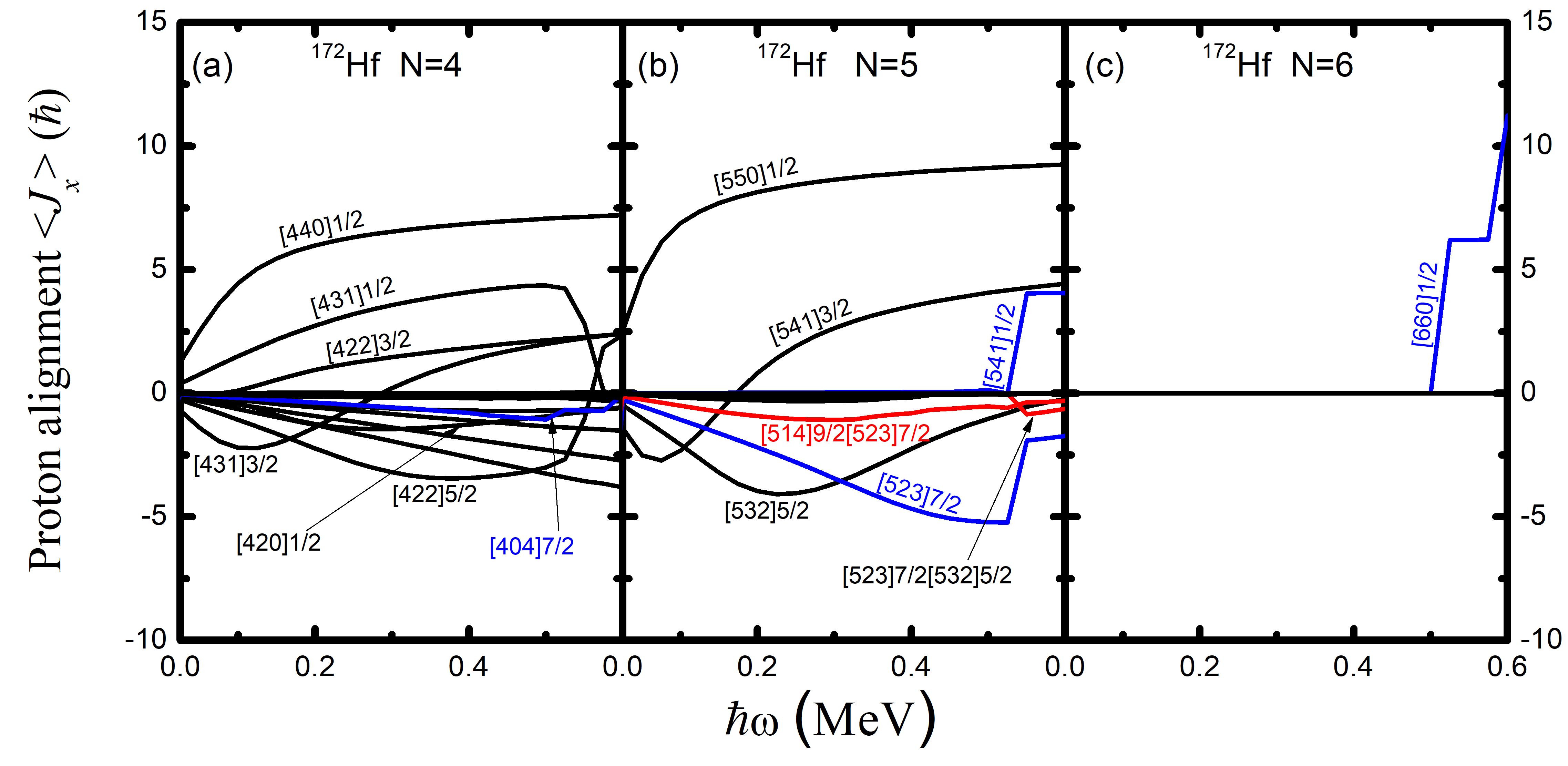}
      \caption{ (Color online) Contributions of each proton orbital in the $N=4,5,6$ major shells to the angular momentum alignment $\langle J_x \rangle$ for the GSB in $^{172}$Hf. The important diagonal (off-diagonal) part $j^{(1)}_{\mu}$ $[ j^{(1)}_{\mu\nu}]$, which is simplified by $\mu$ $(\mu\nu)$, is denoted by blue (red) lines. The orbitals that have little contribution to the upbending (some of these orbitals contribute to the steady increase of the alignment) are denoted by black lines.}
\label{fig:Fig6}
\end{figure}
%%%%%%%%%%%%%%%%%%%%%%%%%%%%%%%%%%%%%%%%%%%%%%%%%%%%%

As analysed above, the upbendings at high rotational frequency $\hbar\omega\approx0.5$ MeV due to the alignment of protons are common for the hafnium isotopes. Here we take $^{172}$Hf for example to discuss. It is well known that the upbending is due to the suddenly increased alignment which are caused by the crossing of the GSB with a pair-broken band based on high-$j$ intruder orbitals. In Fig.~\ref{fig:Fig4}, one can see that the occupation probability of orbitals just below the Fermi surface, $\pi7/2^{+}[404]$ and $\pi7/2^{-}[523]$ decrease to $n_\mu\approx1$ and orbitals above the Fermi surface, $\pi1/2^{+}[660]$ and $\pi1/2^{+}[541]$ increase from $n_\mu=0$ to $n_\mu=1$ at $\hbar\omega=0.5$ MeV. Furthermore, high-$j$ low-$\Omega$ orbital $\pi1/2^{+}[660]$ gets full occupied $(n_\mu\approx2)$ and the low-$j$ high-$\Omega$ orbital $\pi7/2^{+}[404]$ becomes empty $(n_\mu\approx0)$ at frequency $\hbar\omega\geqslant0.55$ MeV.    

The contribution of each proton major shell to the angular momentum alignment $\langle J_x \rangle$ for the GSB in $^{172}$Hf are shown in Fig.~\ref{fig:Fig5}. It shows that the sudden increase of the alignment at $\hbar\omega\approx0.5$ MeV comes from the simultaneous contributions from the proton $N=5$ and $6$ major shells. Furthermore, it is the diagonal part $\sum_{\mu}j^{(1)}_{\mu}$ in Eq.~\ref{eq:J(1)} from the proton $N=5,6$ shells contributes mainly to the sharp raise of the alignment at frequency $\hbar\omega\approx0.5$ MeV. More detailed information about the contribution of each proton orbital in the $N =4,5,6$ major shells to the angular momentum alignments $\langle J_x \rangle$ is presented in Fig.~\ref{fig:Fig6}, where direct term $j_{x}(\mu)$ is simplified by $\mu$ and the interference term $j_{x}(\mu\nu)$ between $\mu$ and $\nu$ is by $\mu\nu$. It shows that the diagonal terms of $j_x(\pi7/2^{-}[523])$ and $j_x(\pi1/2^{-}[541])$ in $N=5$ major shell and $j_x(\pi1/2^{+}[660])$ in $N=6$ major shell contribute mainly to the upbending at $\hbar\omega\approx0.5$ MeV of the GSB in $^{172}$Hf. 

\subsection{Effects of neutron deformed shells and high-$j$ orbitals on the first upbendings}

%%%%%%%%%%%%%%%%%%%%%%%%%%%%%%%%%%%%%%%%%%%%%%%%%%%%%
\begin{figure}[ht] 
\centering
    \includegraphics[width=16cm]{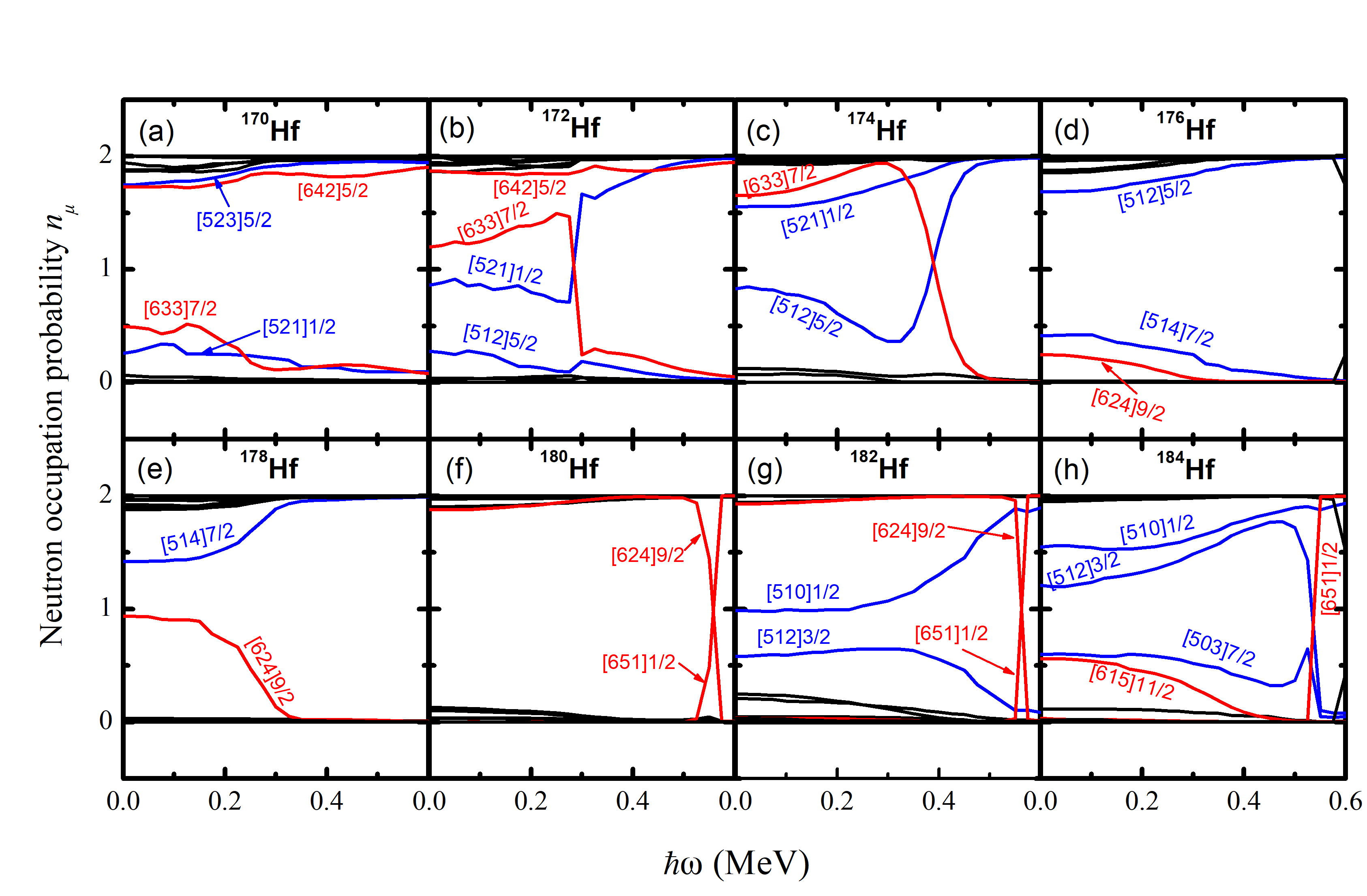}
      \caption{(Color online) Occupation probabilities $n_{\mu}$ of each neutron orbital $\mu$ (including both $\alpha=\pm1/2$) near the Fermi surface for the GSBs in $^{170-184}$Hf. The Nilsson levels far above ($n_{\mu}\approx0$) and far below ($n_{\mu}\approx2$) the Fermi surface are not labelled.}
\label{fig:Fig7}
\end{figure}
%%%%%%%%%%%%%%%%%%%%%%%%%%%%%%%%%%%%%%%%%%%%%%%%%%%%%

It is known from Fig.~\ref{fig:Fig3} that the first upbending of the GSBs is mainly due to the neutron alignments in the even-even isoropes $^{170-184}$Hf. Figure~\ref{fig:Fig7} shows the occupation probability $n_\mu$ (including both signature $\alpha=\pm1/2$) of each neutron orbital $\mu$ near the Fermi surface for the GSBs in $^{170-184}$Hf. The Nilsson levels far above $(n_\mu\approx0)$ and far below $(n_\mu\approx2)$ the Fermi surface are not labelled. It can be seen that the occupation probabilities $n_\mu$ for the GSBs in $^{170}$Hf, $^{176}$Hf and $^{180}$Hf are comparatively pure, i.e. either almost empty $(n_\mu\approx0)$ or fully occupied $(n_\mu\approx2)$. This is because the Fermi surfaces of these three nuclei are just at the deformed neutron shells of $N=98$, $104$ and $108$, respectively. For $^{172}$Hf, $^{174}$Hf, $^{178}$Hf, $^{182}$Hf and $^{184}$Hf, their Fermi surfaces locate between the deformed neutron shells, orbitals above and below the Fermi surface are closed and they are partly occupied due to the pairing correlations. The band-crossings with pair-broken bands based on high-$j$ orbitals $\nu7/3^{+}[633]$ and $\nu1/2^{+}[651]$ happen for the GSBs of $^{172,174}$Hf and $^{180,182,184}$Hf, respectively. This is coincident with the single-particle level structure shown in Fig.~\ref{fig:Fig2} (b) and (c).

 %%%%%%%%%%%%%%%%%%%%%%%%%%%%%%%%%%%%%%%%%%%%%%%%%%%%%
\begin{figure}[ht] 
\centering
    \includegraphics[width=16cm]{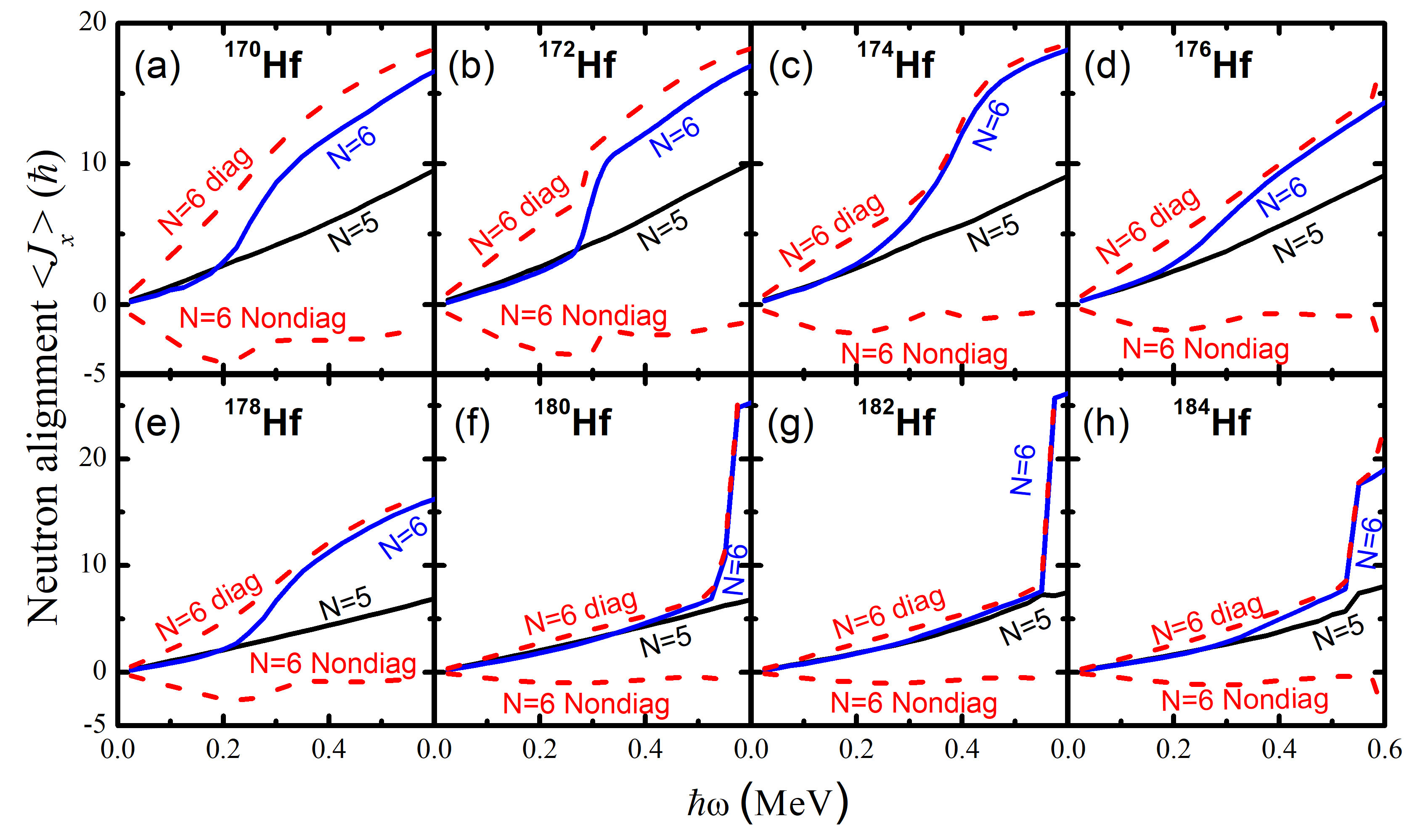}
      \caption{(Color online) Contributions to the angular momentum alignment $\langle J_x \rangle$ from neutron $N=5,6$ major shells for the GSBs in $^{170-184}$Hf. The diagonal $\sum_{\mu}j^{(1)}_{\mu}$ and off-diagonal $\sum_{\mu<\nu} j^{(1)}_{\mu\nu}$ parts in Eq.~\ref{eq:Jx} from the neutron N=6 major shell are shown by red dashed lines.}
\label{fig:Fig8}
\end{figure}
%%%%%%%%%%%%%%%%%%%%%%%%%%%%%%%%%%%%%%%%%%%%%%%%%%%%%

The contribution of each neutron major shell to the angular momentum alignment $\langle J_x \rangle$ is shown in Fig.~\ref{fig:Fig8}. It is seen that the sudden raise of the alignment is due to the contribution of the neutron $N=6$ major shell for all the hafnium isotopes $^{170-184}$Hf. More detailed, the suddenly gained alignments of GSBs in $^{172,174}$Hf come from the contributions of both diagonal $\sum_{\mu}j^{(1)}_{\mu}$ and off-diagonal $\sum_{\mu<\nu} j^{(1)}_{\mu\nu}$ parts of the neutron $N=6$ shell, that in $^{170,176,178}$Hf are from the off-diagonal part and in $^{180-184}$Hf from the diagonal part. One can see that besides the diagonal part, the off-diagonal part also plays an very important role to nuclear rotational properties.

%%%%%%%%%%%%%%%%%%%%%%%%%%%%%%%%%%%%%%%%%%%%%%%%%%%%%
\begin{figure}[ht] 
\centering
    \includegraphics[width=16cm]{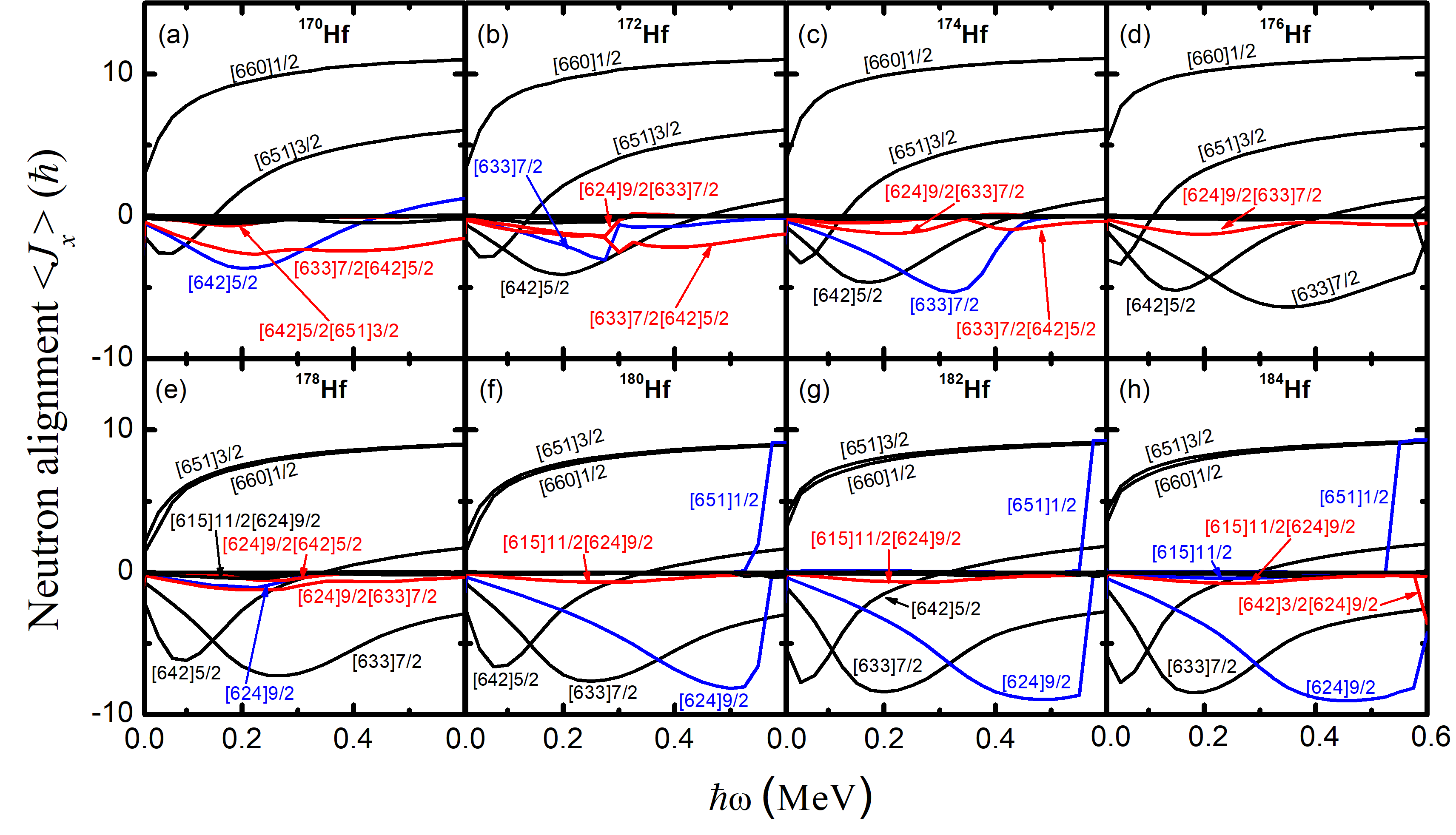}
      \caption{(Color online) Contributions of each neutron orbital in the $N=6$ major shell to the angular momentum alignments $\langle J_x \rangle$ for the GSBs in $^{170-184}$Hf. The important diagonal (off-diagonal) part $j^{(1)}_{\mu}$ $[ j^{(1)}_{\mu\nu}]$, simplified by $\mu (\mu\nu)$, is denoted by blue (red) lines. The orbitals that have little contribution to the upbending (some of these orbitals contribute to the steady increase of the alignment) are denoted by black lines.}
\label{fig:Fig9}
\end{figure}
%%%%%%%%%%%%%%%%%%%%%%%%%%%%%%%%%%%%%%%%%%%%%%%%%%%%%

The contribution of each neutron orbital to the angular momentum alignment is presented in Fig.~\ref{fig:Fig9}. For $^{170}$Hf, the Fermi surface just locates at the deformed shell $N=98$. The orbital just below the Fermi surface is $\nu5/2^{-}[523]$. The high-$j$ orbital $\nu5/2^{+}[642]$ rises up as frequency increases and crosses with $\nu5/2^{-}[523]$ at $\hbar\omega=0.15$ MeV and then it becomes the one just below the Fermi surface [see Fig.~\ref{fig:Fig2} (b)]. From Fig.~\ref{fig:Fig9} (a) one can see that the direct term $j_{x}(5/2^{+}[642])$ and the interference terms $j_{x}(\nu5/2^{+}[642]\otimes\nu3/2^{+}[651])$ and  $j_{x}(\nu5/2^{+}[642]\otimes\nu7/2^{+}[633])$ at $\hbar\omega\approx0.2$ MeV contribute mainly to the sudden raise of the neutron $N=6$ alignment. 

$^{172,174}$Hf are the two nuclei which locate between the two deformed neutron shells $N=98,104$. It can be seen in Fig.~\ref{fig:Fig2} (b) that the high-$j$ orbital $\nu7/2^{+}[633]$ is just below the Fermi surface at very low frequency and rises up quickly as the frequency increases. The $\nu7/2^{+}[633]$ orbital crosses with $\nu1/2^-[521](\alpha=+1/2)$ at $\hbar\omega\approx0.25$ MeV and with $\nu5/2^-[512]$ at $\hbar\omega\approx0.35$ MeV, respectively. In Fig.~\ref{fig:Fig7} (b), the occupation probabilities $n_\mu$ of orbitals $\nu7/2^{+}[633]$ and $\nu1/2^-[521]$ are partly occupied due to the pairing correlation at the low frequency. At $\hbar\omega\approx0.35$ MeV, $n_{\mu}$ of $\nu7/2^{+}[633]$ and $\nu1/2^-[521]$ orbitals is exchanged, and then the $\nu1/2^-[521]$ orbital is almost fully occupied while $\nu7/2^{+}[633]$ is almost empty at the high frequency region. Figure~\ref{fig:Fig9} (b) shows that the contribution of the diagonal part to the upbending of the GSB at $\hbar\omega\approx0.3$ MeV is mainly due to $j_x(\nu7/2^{+}[633])$ and the off-diagonal parts are due to $j_x(\nu7/2^{+}[633]\otimes\nu9/2^+[624])$ and $j_x(\nu7/2^{+}[633]\otimes\nu5/2^+[642])$ for $^{172}$Hf. As for $^{174}$Hf, the situation is similar, except that the orbital crossed with $\nu7/2^{+}[633]$ is $\nu5/2^-[512]$, and the crossing frequency is $\hbar\omega\approx0.4$ MeV. Since neither $\nu1/2^-[521]$ nor $\nu5/2^-[512]$ is the neutron high-$j$ orbital, occupation of these two orbitals gives little influence to the angular momentum alignment. The contributions to the suddenly increased alignment of GSB in $^{172,174}$Hf come mainly from the alignment of $\nu7/2^{+}[633]$ including both the diagonal part $j_x(\nu7/2^{+}[633])$ and the off-diagonal parts $j_x(\nu7/2^{+}[633]\otimes\nu5/2^+[642])$ and $j_x(\nu7/2^{+}[633]\otimes\nu9/2^+[624])$.  

For $^{176}$Hf, it locates just at the deformed shell $N=104$. Therefore, the occupation probabilities $n_\mu$ is comparatively pure and there is no band-crossing appeared [see the Fig.~\ref{fig:Fig7} (d)]. The $J^{(1)}$ increases gradually with frequency and a very gentle upbending appears around $\hbar\omega=0.3$ MeV, which is mainly due to the contribution of the interference term of the alignment $j_x(7/2^{+}[633]\otimes9/2^+[624])$ [see Fig.~\ref{fig:Fig9} (d)]. 

For $^{178}$Hf, from Fig.~\ref{fig:Fig2} (c), we see that the $\nu7/2^-[514]$ and $\nu9/2^+[624]$ orbitals, the orbitals just below and above the Fermi surface, locate close to each other at the low frequency region. They are partly occupied at frequency $\hbar\omega<0.3$ MeV [see Fig.~\ref{fig:Fig7} (e)]. At higher frequency, the occupation probability becomes quite pure. The upbending of GSB at frequency $\hbar\omega\approx0.3$ MeV is mainly due to the contribution of the high-$j$ intruder orbital $9/2^+[624]$. It can be seen more clearly from Fig.~\ref{fig:Fig9} (e) that the gentle upbending at $\hbar\omega\approx0.3$ MeV comes mainly from the contribution to the alignment of the direct term $j_x(\nu9/2^+[624]$ and interference terms of $j_x(\nu9/2^+[624]\otimes\nu7/2^+[633])$ and $j_x(\nu9/2^+[624]\otimes\nu5/2^+[642])$. 

The rotational behavior of the neutron-richer nuclei $^{180,182,184}$Hf is quite different from the lighter isotopes. The $J^{(1)}$ increases slightly and smoothly with increasing frequency and there is no upbending appeared during the experimental observed frequency range. Cranked Nilsson levels in Fig.~\ref{fig:Fig2} (b), which is obtained by using the quadrupole deformation $\varepsilon_{2}=0.254$, show that $\nu1/2^+[651]$ decreases rapidly with increasing frequency and crosses with $\nu9/2^+[624]$, $\nu1/2^-[510]$ and $\nu3/2^-[512]$ at $\hbar\omega\approx0.4$, $\approx0.15$ and $\approx0.1$ MeV, respectively. The $\nu1/2^+[651]$ orbital is a high-$j$ low-$\Omega$ orbital. Band-crossing with a high-$j$ low-$\Omega$ orbital will lead to a sharp backbending/upbending of a rotational band. However, $J^{(1)}$s of the GSB in $^{180,182,184}$Hf vary smoothly at the observed frequency region. As shown in Fig.~2d in Ref.~\cite{NilssonS1969_NPA131}, the high-$j$ intruder orbital $\nu1/2^+[651]$ goes down very quickly as the quadrupole deformation increases. Therefore, to avoid the influence of the $\nu1/2^+[651]$ orbital, smaller quadrupole deformation is adopted to calculate the GSBs of $^{180,182,184}$Hf, and the single-particle levels are obtained as shown in Fig.~\ref{fig:Fig2} (c). Based on such single-particle level structure, upbendings at $\hbar\omega\approx0.5$ MeV are predicted by the theoretical calculations for the GSBs in $^{180,182,184}$Hf, which attribute to the simultaneous sharp increased proton and neutron alignments [See in Fig.~\ref{fig:Fig3} (f-h)]. The proton contribution is simliar to the situation of $^{172}$Hf discussed above. For neutrons, as shown in Fig.~\ref{fig:Fig7} (f-h) that band-crossing between the band based on the $\nu1/2^+[651]$ orbital and GSBs occurs at $\hbar\omega\approx0.5$ MeV. The suddenly increased neutron alignment comes mainly from contribution of the diagonal part in neutron $N=6$ major shell. Figure~\ref{fig:Fig9} (f-h) shows that the alignments gained after the band-crossing frequency are from contributions of the direct terms $j_{x}(\nu1/2^+[651])$ and $j_{x}(\nu9/2^+[624])$. 

\subsection{Effects of neutron pairing correlations on band-crossings and backbendings}

%%%%%%%%%%%%%%%%%%%%%%%%%%%%%%%%%%%%%%%%%%%%%%%%%%%%%
\begin{figure}[ht] 
\centering
    \includegraphics[width=16cm]{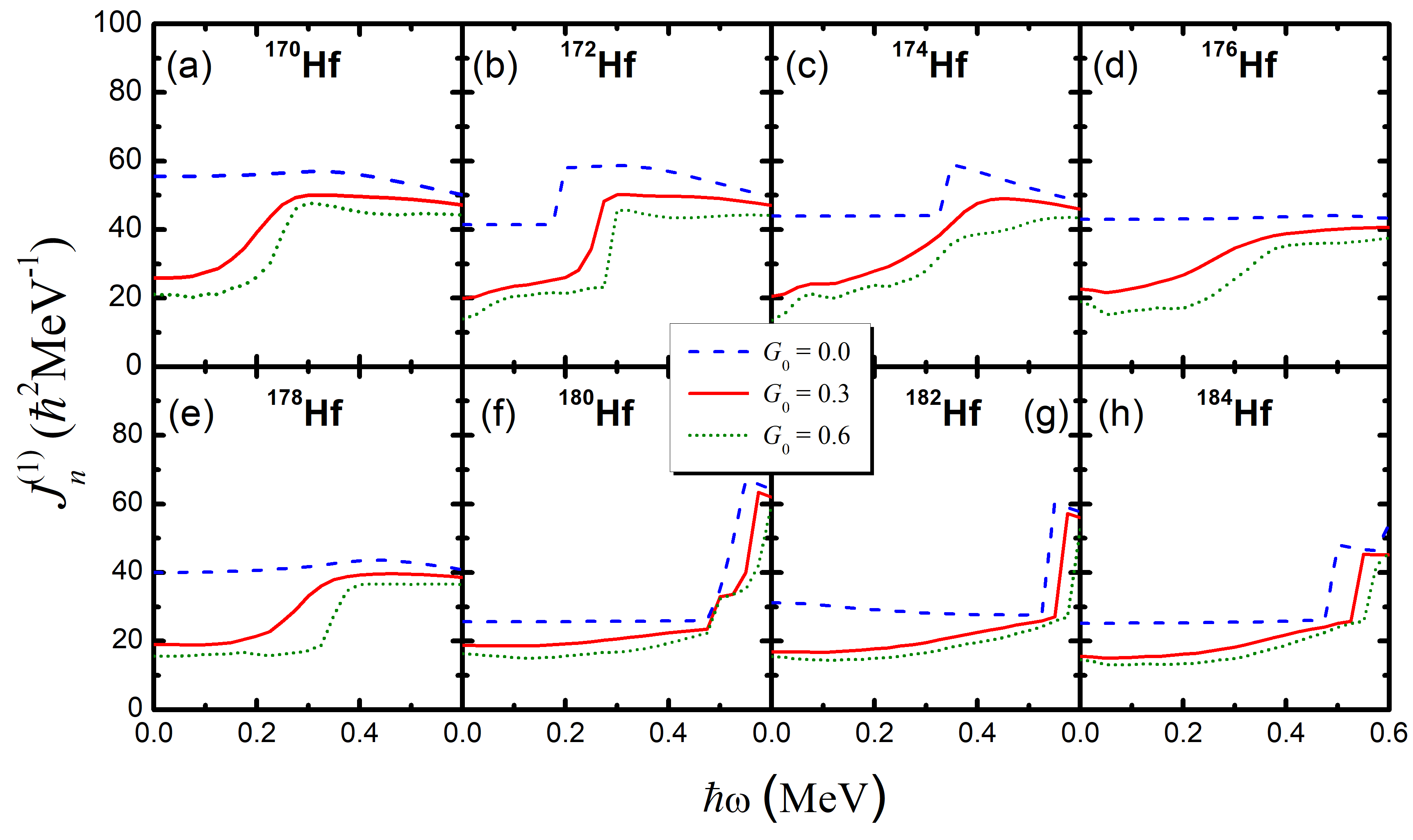}
      \caption{(Color online) The neutron kinematic moment of inertia $J^{(1)}_n$ calculated by using the monopole effective pairing strengths $G_0=0.0$ (blue dashed lines), $G_0=0.3$ (red solid lines) and $G_0=0.6$ (olive dotted lines).}
\label{fig:Fig10}
\end{figure}
%%%%%%%%%%%%%%%%%%%%%%%%%%%%%%%%%%%%%%%%%%%%%%%%%%%%%

%%%%%%%%%%%%%%%%%%%%%%%%%%%%%%%%%%%%%%%%%%%%%%%%%%%%%
\begin{figure}[ht] 
\centering
    \includegraphics[width=16cm]{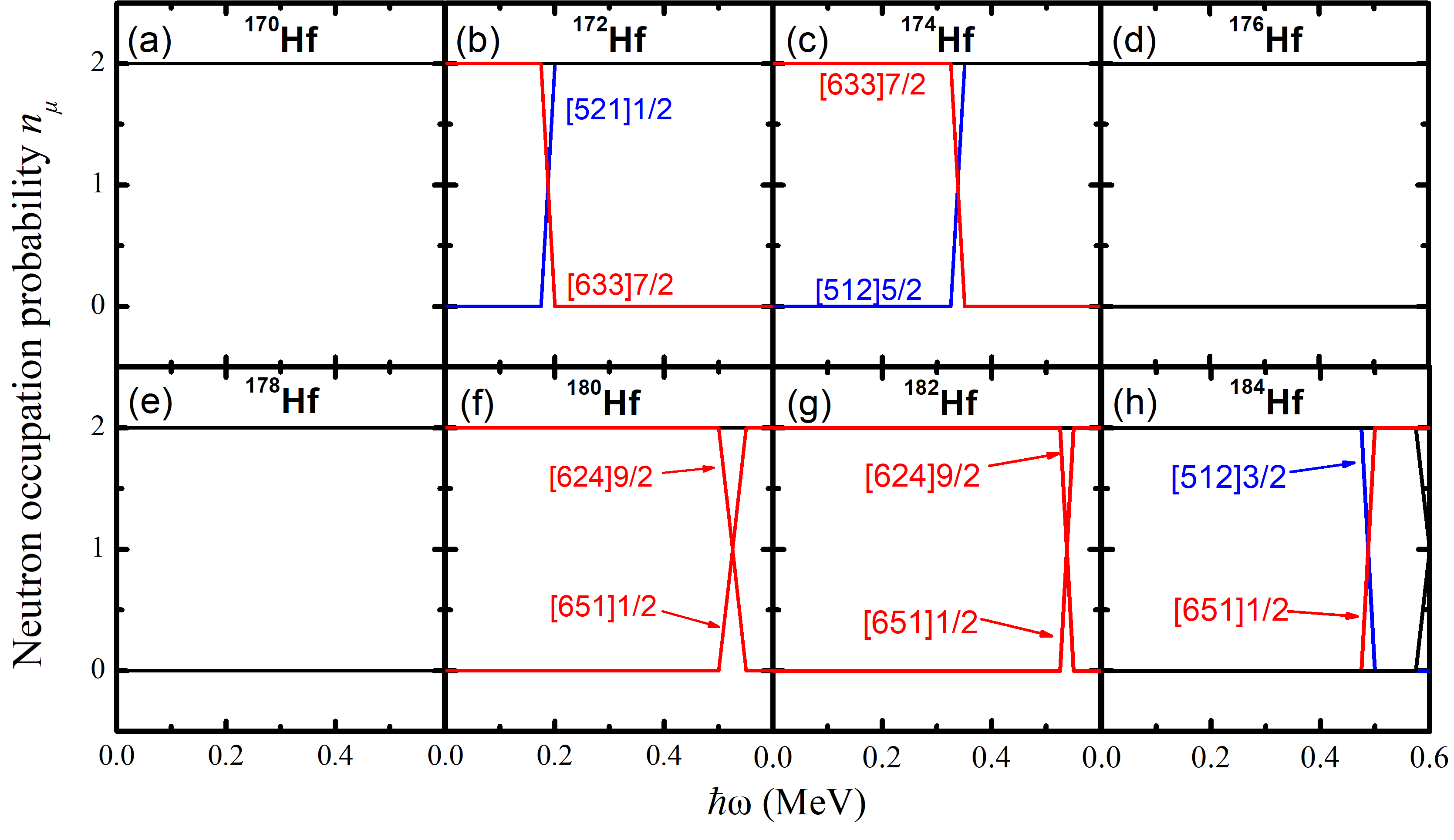}
      \caption{(Color online) The same as Fig.~\ref{fig:Fig7}, but for the results without pairing correlations.}
\label{fig:Fig11}
\end{figure}
%%%%%%%%%%%%%%%%%%%%%%%%%%%%%%%%%%%%%%%%%%%%%%%%%%%%%

%%%%%%%%%%%%%%%%%%%%%%%%%%%%%%%%%%%%%%%%%%%%%%%%%%%%%
\begin{figure}[ht] 
\centering
    \includegraphics[width=16cm]{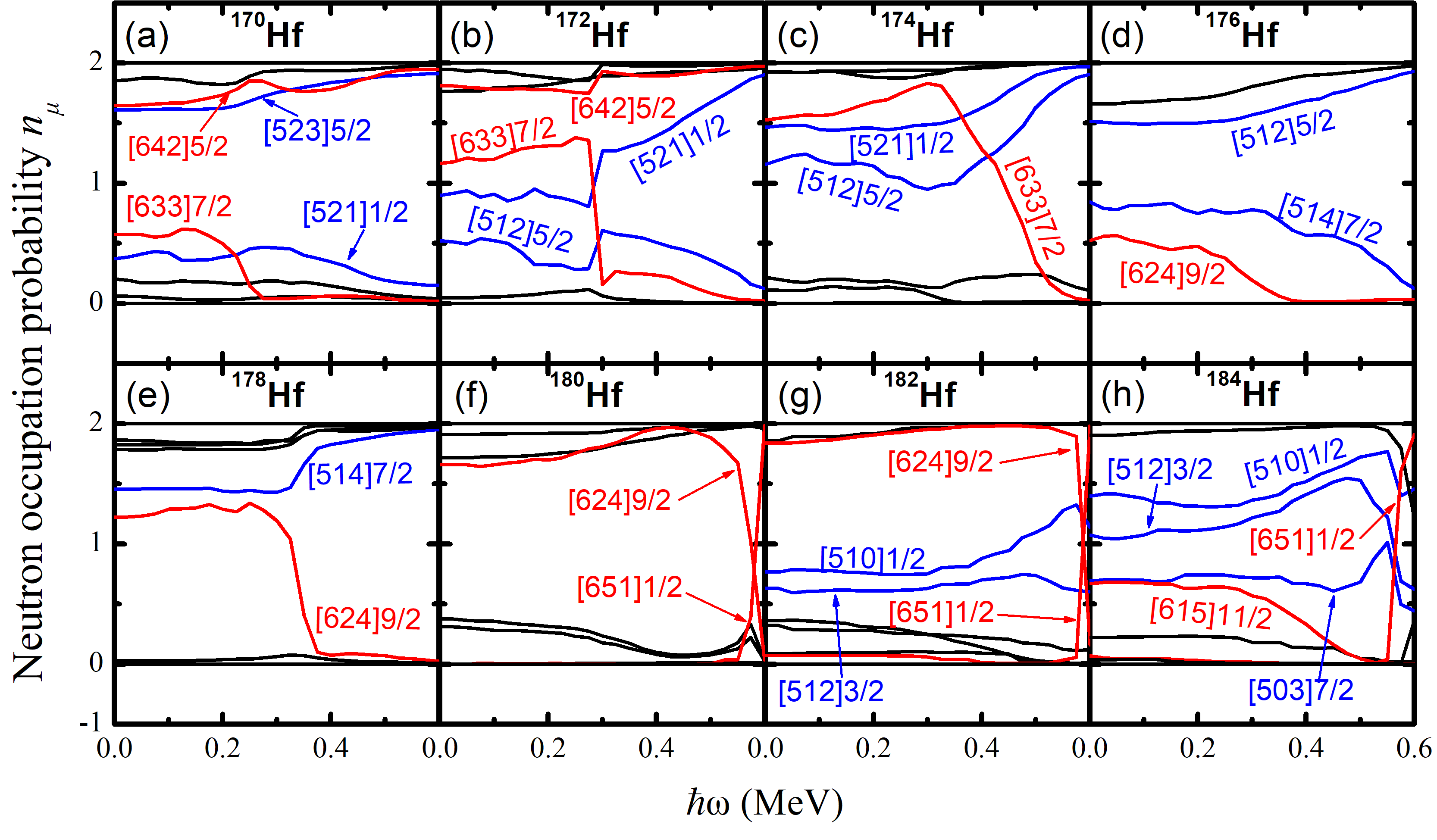}
      \caption{(Color online) The same as Fig.~\ref{fig:Fig7}, but for the results obtained by using the effective pairing strength $G_0=0.6$.}
\label{fig:Fig12}
\end{figure}
%%%%%%%%%%%%%%%%%%%%%%%%%%%%%%%%%%%%%%%%%%%%%%%%%%%%%

From the above discussion, one can see that deformation, high-$j$ intruder orbitals and deformed neutron shell gaps influence the rotational properties and upbendings strongly. It will be interesting to investigate, with deformation and single-particle level structure unchanged, the influence of pairing on the upbendings in $^{170-184}$Hf. 

Figure~\ref{fig:Fig10} shows the neutron kinematic moment of inertia $J^{(1)}_n$ calculated by using the monopole effective pairing strengths $G_0=0.0$ (blue dashed lines), $G_0=0.3$ (red solid lines) and $G_0=0.6$ (olive dotted lines). When the theoretical calculations performed without pairing correlations, $J^{(1)}_n$ keep almost constant during the whole rotational frequency for $^{170,176,178}$Hf while obvious upbendings are presented at $\hbar\omega\approx0.2$, 0.35 and 0.5 MeV for $^{172}$Hf, $^{174}$Hf and $^{180,182,184}$Hf, respectively. These upbendings are caused by the band-crossings shown in Fig.~\ref{fig:Fig11} where the occupation probabilities $n_{\mu}$ of each neutron orbital $\mu$ near the Fermi surface are obtained without pairing correlations. As shown in Fig.~\ref{fig:Fig11} that the occupation probability is either equal to 2 or equal to 0 when there is no pairing, and the band-crossing frequencies for the GSBs in $^{172,174,180,182,184}$Hf are the same with the crossing of the single-particle levels presented in Fig.~\ref{fig:Fig2}. 

When calculations include pairing correlations, besides what mentioned above in $^{172,174,180,182,184}$Hf, upbendings appear also in the GSBs of $^{170,176,178}$Hf [see Fig.~\ref{fig:Fig11} (a) (d) and (e)]. Due to the pairing, some of the configurations are mixed and orbitals below and above the Fermi surface become partial occupied as shown in Fig.~\ref{fig:Fig7} (a) (d) and (e). Note that the results in Fig.~\ref{fig:Fig7} are obtained with $G_0=0.3$. It gets more pronounced with stronger pairing correlation, like the results calculated with pairing strength $G_0=0.6$ presented Fig.~\ref{fig:Fig12} (a) (d) and (e). For the bandhead of the GSBs, the results with pairing is lower than the results without pairing for all the isotopes studied here. Nevertheless, after band crossing, due to the Coriolis anti-pairing force, the results with and without pairing get very close to each other as shown in Fig.~\ref{fig:Fig10}. 

In the most recent work of Miller \textit{et al}.~\cite{MillerS2019_PRC100_14302}, the band-crossing frequencies versus pairing were investigated for $^{166}$Hf. The calculation leads to a rise of crossing frequency with larger pairing (see Fig.~7 in Ref.~\cite{MillerS2019_PRC100_14302}). Same conclusion can be seen in Fig.~\ref{fig:Fig10} for heavier hafnium isotopes $^{170-178,182-184}$Hf. Upbendings are delayed to higher frequency when larger pairing correlation strengths are used. However, $^{180}$Hf is out of the line somehow. Besides, we notice from Fig.~\ref{fig:Fig10} that the influence of pairing is less striking for $^{180,182,184}$Hf. This indicates that the deformed neutron shell $N=108$ is more pronounced.
 
\section{summary}

The GSBs in the even-even hafnium isotopes $^{170-184}$Hf are investigated systematically by using the cranked shell model with pairing correlations treated by the particle-number conserving method. The experimental kinematic moments of inertia $J^{(1)}$ are reproduced very well by theoretical calculations. The microscopic mechanism of the variation of $J^{(1)}$ versus rotational frequency is explained by the band-crossing of a pair-broken band based on high-$j$ intruder orbitals with the GSB. The different behavior of the GSB between the lighter isotopes $^{170-178}$Hf and the heavier isotopes $^{180-184}$Hf is revealed through the alignment of high-$j$ intruder orbitals and the neutron deformed shells of $N=98,104,108$ and $N=116$.  
 
The observed upbending at the high frequency $\hbar\omega\approx0.5$ MeV of GSB in $^{172}$Hf is due to the crossing of the pair-broken band based on the proton high-$j$ orbitals $\pi1i_{13/2}$ $(1/2^{+}[660])$, $\pi1h_{9/2}$ $(1/2^{-}[541])$ and orbital $\pi1h_{11/2}$ $(7/2^{-}[523])$ with the GSB. Since the proton Fermi-surfaces are the same, the proton alignment at $\hbar\omega\approx0.5$ MeV is common for all the hafnium isotopes $^{170-184}$Hf. The upbendings, which result from the proton alignment, are predicted at $\hbar\omega\approx0.5$ MeV for the GSB of the even-even hafnium isotopes $^{170,174-178}$Hf. 

For the lighter even-even isotopes $^{170-178}$Hf, besides the upbending resulted from the proton alignment at the high frequency, upbendings of the GSBs appear at low frequency $\hbar\omega=0.2-0.3$ MeV. These upbendings attribute to the alignment of the neutron high-$j$ orbital $\nu1i_{13/2}$. The deformed neutron shells are presented at $N=98,104,108$ for $^{170-178}$Hf. The upbending at low frequency of GSB in the isotope below the N=98 deformed shell, i.e. $^{170}$Hf, is due to the contribution from the alignment of the neutron orbital $\nu1i_{13/2}$ ($5/2^{+}[642]$), upbendings in the isotopes between the N=98,104 deformed shells, i.e. $^{172,174}$Hf, attribute to the sudden alignment of the neutron orbital $\nu1i_{13/2}$ ($7/2^{+}[633]$), upbendings in the isotopes between the N=104,108 deformed shells, i. e. $^{176,178}$Hf, mainly result from the alignment of the orbital $\nu1i_{13/2}$ ($9/2^{+}[624]$). 

For the heavier even-even isotopes $^{180-184}$Hf, there is no obvious nucleon alignment of GSBs was observed at the low frequency in the experiment. To reproduce the experimental data, compared to that used for the lighter isotopes, smaller quadrupole deformation parameters are adopted for $^{180,182,184}$Hf. With smaller deformation, in addition to $N=98,104,108$, the $N=116$ deformed neutron shell emerges and the high-$j$ intruder orbital $\nu2g_{9/2}$ $(1/2^{+}[651])$ rises to higher position in the single-particle levels. This leads to a delay of the band-crossing of the pairing-broken band based on the $\nu1/2^{+}[651]$ orbital with the GSB in $^{180,182,184}$Hf. In the PNC-CSM calculation, upbendings are predicted at the high frequency $\hbar\omega\approx0.5$ MeV for the GSB in the even-even isotopes $^{180-184}$Hf, which are due to the suddenly simultaneous alignments of the neutron high-$j$ intruder orbital $\nu2g_{9/2}$ ($1/2^{+}[651]$) and the proton $\pi1i_{13/2}$, $\pi1h_{9/2}$ orbitals. 

The pairing correlation plays a very important role in the rotational properties of GSBs in even-even isotopes $^{180-184}$Hf. The upbendings at frequency $\hbar\omega\approx0.2$ in the GSBs of $^{170,176,178}$Hf arise due to the pairing correlations. The upbending and band-crossing frequencies increase with larger pairing correlation strength generally. Among the deformed neutron shells mentioned above, $N=108$ is the most pronounced one. 

\section*{Acknowledgments}
This work is supported by the National Natural Science Foundation of China (Grant Nos. 11775112 and 11275098) and the Priority Academic Program Development of Jiangsu Higher Education Institutions.

\section*{References}
\bibliographystyle{apsrev4-1}
\bibliography{../../../References/ReferencesXT}
%\bibliography{library}

\end{document}